\documentclass[12pt]{article}

\begin{document}

\thispagestyle{empty}
\renewcommand{\thefootnote}{\fnsymbol{footnote}}

\begin{flushright}
{\small
SLAC--PUB--8313\\
December 1999\\}
\end{flushright}

\vspace{.8cm}

\begin{center}
{\bf\large
NLC Extraction Line Studies\footnote{Work supported by
Department of Energy contract DE--AC03--76SF00515.}}

\vspace{1cm}

Y.~M.~Nosochkov and T.~O.~Raubenheimer\\
Stanford Linear Accelerator Center, Stanford University,
Stanford, CA  94309\\
\end{center}

\vfill

\begin{center}
{\bf\large
Abstract }
\end{center}

\begin{quote}
In this note, we briefly review the current lattice of the NLC
extraction line which was designed for the nominal NLC beam
parameters~\cite{ext:lum-nom,ext:param}.  Then we describe the beam
parameters for the high luminosity option with larger beam disruption
parameter~\cite{ext:lum-high} and discuss its effect on beam loss in
the extraction line.  Finally, we present a summary of the optics study
aimed at minimizing the beam loss with high disruption beams.
\end{quote}

\vfill

\newpage

\pagestyle{plain}

\section{Introduction}

In a linear collider, the rate of colliding bunches tends to be low 
when compared with a colliding beam storage ring.  
Fortunately, because the beams are not stored,
the disruptive effect of the beam-beam forces is not a dynamical limitation
and thus the beams can be focused to very small spot sizes to attain
the high charge densities necessary to achieve the desired luminosity.  When
the high energy electron and positron beams collide, the space charge 
forces of the opposing beam cause large deflections.  
This has two effects: first, the effective
beam emittance of the outgoing beam is increased due to the 
non-linearity of the beam-beam force and, second, when the high energy
particles are deflected, they can radiate a significant fraction of their
energies as synchrotron radiation which is referred to as beamstrahlung.

The NLC extraction line must be designed to transport this disrupted 
particle beam and the beamstrahlung photons away from the interaction 
point (IP) to dumps.  In addition, the extraction line should 
provide diagnostics to fully instrument the
collisions.  This includes the ability to measure the outgoing beam angle and
position, measure the energy spectrum of the outgoing beam, and measure the
polarization of the outgoing beam.  Because of the increased angular 
divergence and energy spread of the disrupted beam, it is a very 
difficult task to capture this beam and control it sufficiently well to 
make measurements of its properties.

In the next sections, we will review the present NLC design for the 
extraction line
and some of the studies that have been made to further improve the performance.
Many details of the present design are described in Ref.~\cite{ext:pac99}.

\section{Current Lattice Design}

\begin{figure}[t]
\includegraphics{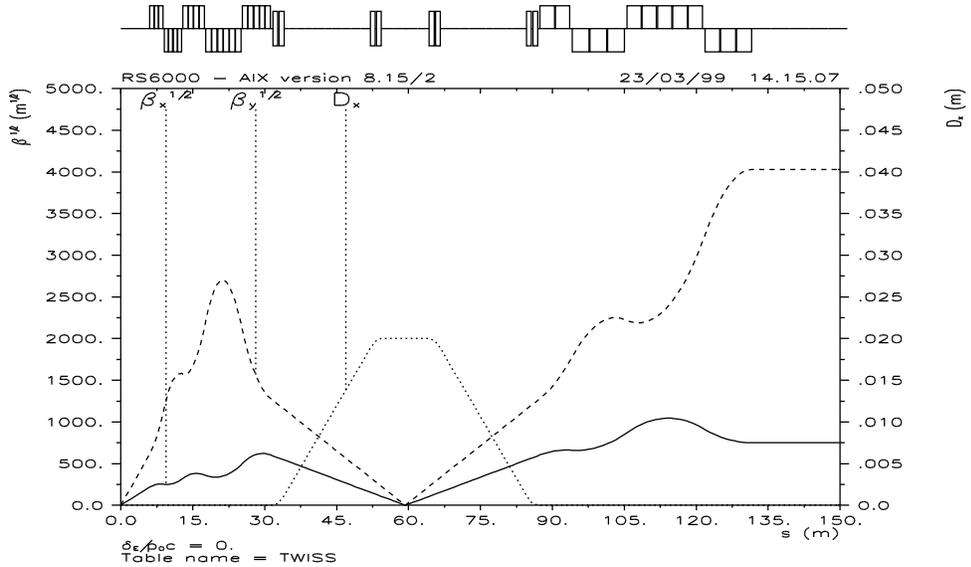}
\vspace{76mm}
\caption{Current lattice of the NLC extraction line (IP~at~$s=0$).}
\label{ext:lat-nom}
\vspace{0mm}
\end{figure}

The current lattice of the NLC extraction line is shown in 
Fig.~\ref{ext:lat-nom}.  The magnet apertures are  
designed to allow the outgoing main beam and beamstrahlung photons to be 
transported to one shared dump at about 150~m from IP.  The design
provides 6~m of free space after IP to avoid interference with the
quadrupoles needed to focus the incoming beam.

The optics consists of two multi-quad 
systems in the beginning and at the end of the beam-line, and a four-bend 
horizontal chicane in the middle.  The first set of quadrupoles focuses the
beam to a waist at the center of the horizontal chicane and then the second
set of quadrupoles creates a parallel beam at the dump.
The chicane creates 2~cm of horizontal dispersion at the secondary 
IP which will facilitate the measurement of the disrupted beam 
energy spread from which one can infer the luminosity spectrum.  
In addition, the chicane separates the particle beam from the 
core of the beamstrahlung photons which will allow measurements of the photon
distribution.  Finally, the electron beam waist and chicane provide an ideal 
location to measure the beam polarization using a Compton laser polarimeter, 
similar to the one developed for the SLC~\cite{ext:polarimeter}. 

\begin{figure}[b!]
\includegraphics{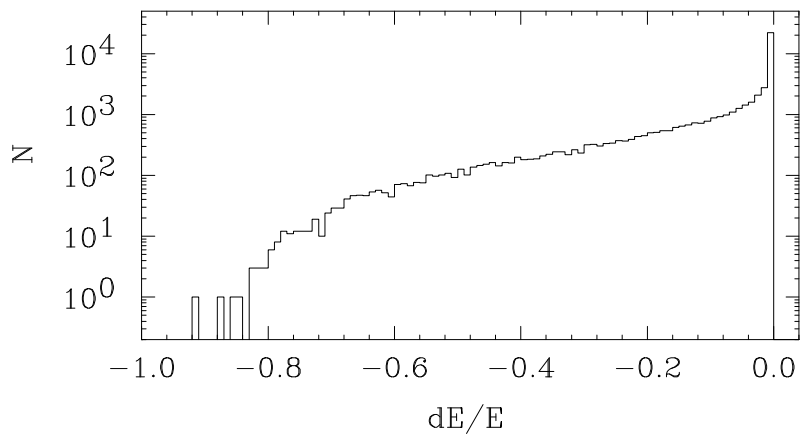}
\vspace{56mm}
\caption{Energy distribution for the nominal disruption beam
(NLC~1~TeV~case~A; 50,000~particles).}
\label{ext:de-nom}
\vspace{0mm}
\end{figure}

One of the main issues for the extraction line design is the
minimization of beam loss; this is necessary to control
backgrounds in the detector as well as the instrumentation in the 
extraction line.  Most of the losses occur for the
very low energy particles which experience large deflections in the 
magnets. For the typical NLC beam parameters~\cite{ext:lum-nom,ext:param}, 
the beam-beam collisions
generate an extremely broad energy spread in the outgoing
beams.  The spectrum at 500 GeV in the center-of-mass (cms) is narrower,
but, with collisions at 1 TeV cms, the spectrum is so wide that 
there is still a significant number of electrons with energy
deviations $\delta=\frac{\Delta p}{p}$ below -80\%~\cite{ext:kathy}. 
An example of the disrupted beam energy distribution is shown in 
Fig.~\ref{ext:de-nom} which corresponds to the beam parameters 
with the largest disruption at an 
energy of 1 TeV cms in the nominal NLC operation; 
these beam parameters are listed in Table~1 and are referred to 
as the `NLC 1~TeV case~A' parameter set. In this paper, 
we will refer to this case as the `nominal disruption' case.

\begin{table}[t]
\vspace{-0mm}
\begin{center}
\caption{IP parameters for the nominal disruption beam (NLC~1~TeV~case~A).}
\medskip
\begin{tabular}{lcc}
\hline
Beam parameter & Undisrupted & Disrupted \\
 & $x/y$ & $x/y$ \\
\hline
Emittance (m$\cdot$rad) [$10^{-13}$] & 39/0.59 & 120/1.02 \\
rms beam size (nm) & 198/2.7 & 198/3.2 \\
rms divergence ($\mu$rad) & 20/22 & 125/33 \\
$\beta^{*}$ (mm) & 10/0.125 & 3.259/0.103 \\
$\alpha^{*}$ & 0/0 & 1.805/0.306 \\
Energy cms (GeV) & \multicolumn{2}{c}{1046} \\
Particles per bunch & \multicolumn{2}{c}{$0.75 \cdot 10^{10}$} \\
Bunches per train & \multicolumn{2}{c}{95} \\
Repetition rate (Hz) & \multicolumn{2}{c}{120} \\
Disruption parameter & \multicolumn{2}{c}{0.094/6.9} \\
Average energy loss per particle & \multicolumn{2}{c}{9.5\%} \\
\hline
\end{tabular}
\label{ext:param-nom}
\end{center}
\vspace{0mm}
\end{table}

The low energy over-focusing in the extraction line is minimized with the 
use of alternating gradient multiple quadrupole systems rather than simple
doublets. The strength of individual quadrupoles in these systems
is reduced compared to the doublet. This results in less over-focusing,
smaller amplitude of the low energy particle oscillations and  
reduced beam loss. 

\begin{figure}[b!]
\includegraphics{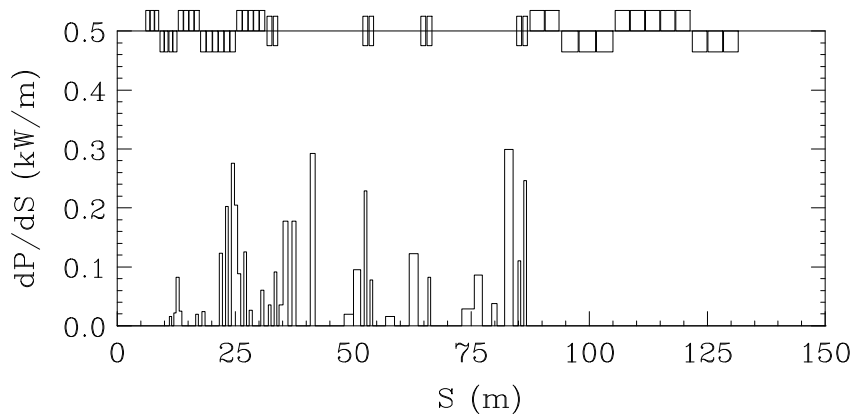}
\vspace{54mm}
\caption{Power loss distribution for the nominal disruption beam
in the current lattice (NLC~1~TeV~case~A).}
\label{ext:power-nom-dump12}
\vspace{0mm}
\end{figure}

\begin{figure}[t]
\includegraphics{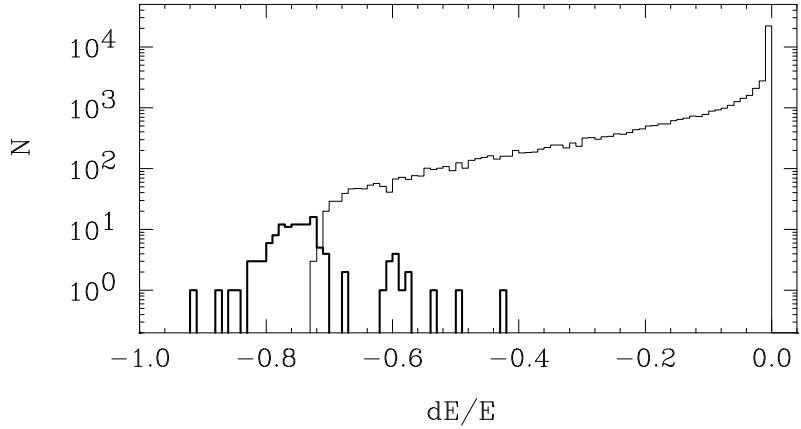}
\vspace{56mm}
\caption{Energy distribution for the lost (thick line) and survived
(thin) nominal disruption beam particles in the current lattice 
(NLC~1~TeV~case~A; 50,000~particles).}
\label{ext:de-lost-nom-dump12}
\vspace{0mm}
\end{figure}

For the beam parameters in Table~1 and optimized quadrupole
strengths, the beam power losses in the extraction line are below 
0.3~kW/m (see Fig.~\ref{ext:power-nom-dump12}) with the total loss of 4.8~kW 
or 0.25\% particles; the energy distribution of the
lost and survived particles is shown in Fig.~\ref{ext:de-lost-nom-dump12}.
It should be noted that at a cms energy of 500 GeV, the losses are roughly 
an order-of-magnitude smaller.

\begin{figure}[t]
\includegraphics{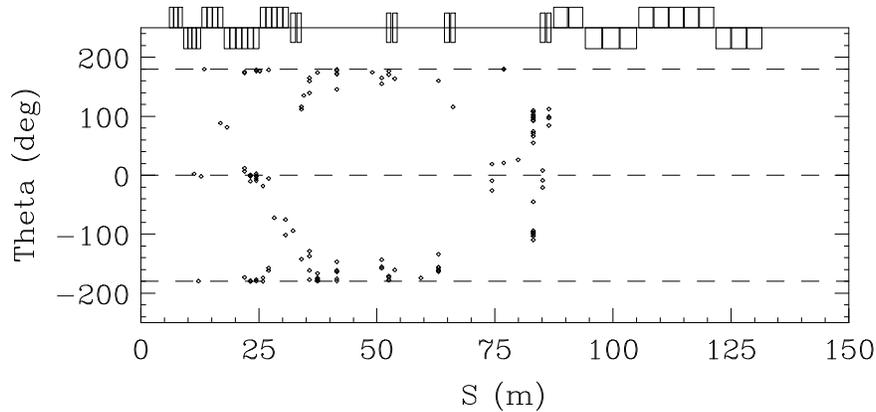}
\vspace{54mm}
\caption{Azimuth distribution of lost particles (dots) for 
the nominal disruption beam (NLC~1~TeV~case~A).}
\label{ext:theta-lost}
\vspace{0mm}
\end{figure}

The azimuth distribution of
particle loss is shown in Fig.~\ref{ext:theta-lost}, where
$\theta=atan \frac{y}{x}$ and $x$, $y$ are the final coordinates of the
lost particles. Due to somewhat larger horizontal beam size in the
first half of the extraction line, there are more losses in the $x$-plane
($\theta=0,\pm 180^{\circ}$) than in the $y$-plane
($\theta=\pm 90^{\circ}$). Though not yet fully studied, it is
hoped that this level of beam loss can be safely
absorbed without significant complications.

\section{High Disruption Beam}

\begin{figure}[t]
\includegraphics{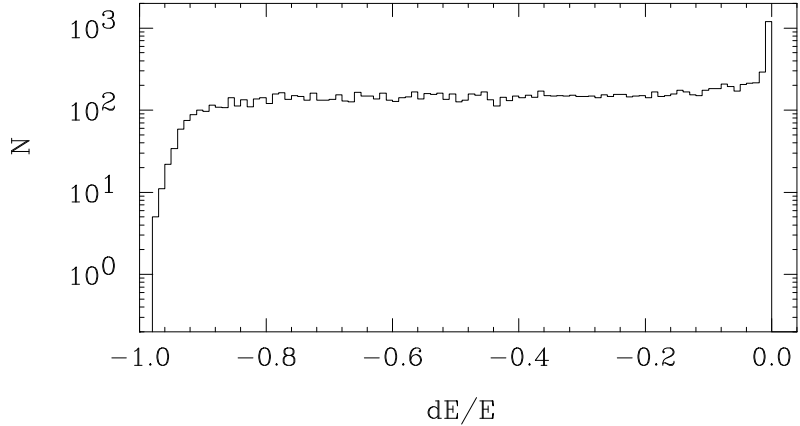}
\vspace{56mm}
\caption{Energy distribution for the high disruption beam
(15,000~particles).}
\label{ext:de-high}
\vspace{0mm}
\end{figure}

It has been reported recently that the NLC luminosity can be
increased several times by using near-equal horizontal and vertical 
beta functions at the IP~\cite{ext:lum-high}. In this scenario,
the undisrupted horizontal beam size at the IP is reduced and the 
vertical size is increased. The latter has an additional advantage of 
much looser tolerances in the final focus system. The disadvantage
of this scenario is a significant increase in the angular beam
divergence and beam energy spread after the collision. The IP
parameters for the `high disruption' case, similar to the 
case in Table~1, are listed
in Table~2~\cite{ext:lum-high}. The corresponding 
energy distribution for the disrupted beam is shown in 
Fig.~\ref{ext:de-high}.

\begin{table}[t]
\vspace{-0mm}
\begin{center}
\caption{IP parameters for the `round(er) beam' collisions at 1~TeV.}
\medskip
\begin{tabular}{lcc}
\hline
Beam parameter & Undisrupted & Disrupted \\
 & $x/y$ & $x/y$ \\
\hline
Emittance (m$\cdot$rad) [$10^{-13}$] & 39/0.59 & 235/7.4 \\
rms beam size (nm) & 62.5/7.7 & 67.3/10.7 \\
rms divergence ($\mu$rad) & 62.5/7.7 & 531/103 \\
$\beta^{*}$ (mm) & 1.0/1.0 & 0.192/0.154 \\
$\alpha^{*}$ & 0/0 & 1.143/1.100 \\
Energy cms (GeV) & \multicolumn{2}{c}{1046} \\
Particles per bunch & \multicolumn{2}{c}{$0.75 \cdot 10^{10}$} \\
Bunches per train & \multicolumn{2}{c}{95} \\
Repetition rate (Hz) & \multicolumn{2}{c}{120} \\
Disruption parameter & \multicolumn{2}{c}{0.85/6.9} \\
Average energy loss per particle & \multicolumn{2}{c}{41\%} \\
\hline
\end{tabular}
\label{ext:param-high}
\end{center}
\vspace{0mm}
\end{table}

Tables~1 and 2 show that the angular divergence of the high
disruption beam is increased by a factor of 3-4 compared to the
nominal beam. This larger divergence translates into larger beam size
in the quadrupoles, and therefore larger quadrupole apertures 
are required. Secondly, as seen in Fig.~\ref{ext:de-nom} and 
\ref{ext:de-high}, the low energy tail in the high disruption
beam is much more populated than in the nominal beam option. For comparison,
the nominal disruption beam has about 1.2\% low energy particles with
$\delta$ below -60\%, while the high disruption beam has 30\% particles 
in this range and 10.6\% particles with $\delta$ below -80\%. 
To achieve beam loss in the high
disruption case similar to the nominal case (about 5~kW), the chromatic 
bandwidth of the extraction line has to be increased to $\delta=-92$\% 
from the -70\% for the nominal beam. Thus, the lowest particle energy
accepted by the optics has to be further reduced by a factor of 4.
Note that at $\delta=-92$\% the particles experience 12 times larger
deflections in the magnets compared to the on-momentum beam.
These very large deflections make it difficult to confine the tail 
particles within the aperture.

\section{Study of High Disruption Beam Loss}

As a first step, we verified the high disruption beam loss for the
current extraction line design. In the simulations, we used
disrupted beam distributions at the IP generated with GUINEA--PIG
code~\cite{ext:kathy}, and then tracked the beam to the dump using
modified DIMAD code which correctly simulates large energy
deviations~\cite{ext:dimad}. With 15,000 initial particles, the
total beam loss was 28.2\% particles, all in the range of $\delta$
below -60\%. Most of these losses are due to the very low
energy particles, but are also enhanced by
the large beam divergence at the IP. About half of the beam loss (15.2\%)
occurred in the five-quadrupole system after IP.

Clearly, the above losses need to be significantly reduced for
a realistic design with high disruption beams. Below
we describe various methods we tried to reduce the beam loss.
However, more studies are needed for an acceptable solution.

Finally, note that in this paper we discuss one particular option for the high
disruption beam parameters (Table~2). There are other scenarios with less
severely disrupted beams which still benefit from the near-equal beta
functions at the IP~\cite{ext:lum-high}.  There are also parameters
for high luminosity operation at 1 TeV with `flat' beams that 
have much smaller disruption and beamstrahlung~\cite{ext:param}.

\subsection{Optics Modifications}

As mentioned earlier, a large portion of the high disruption beam 
(15.2\% particles) is lost in the five quadrupoles after IP. It indicates 
that the optics optimization has to begin with this quadrupole system. The 
main causes of the excessive beam loss are: much larger than nominal 
the low energy tail and increased beam divergence at the IP.

The larger IP divergence increases the beam size in the quadrupoles. 
To improve the quadrupole acceptance, we increased the
quadrupole aperture by a factor of 2 (for instance, from $a_{p}=10$ to 23~mm 
in the first quad, etc., where $a_{p}$ is the pole tip radius). For realistic
design, we kept the quadrupole pole tip field $B_{p}$ below 12~kG. The 
resultant reduction in field gradient $B_{p}/a_{p}$ was compensated by 
lengthening of the quadrupoles. The longer quadrupoles, in turn, 
increase the beam size; thus, both the length and aperture were 
optimized.
 
The quadrupole focusing depends on energy as $\frac{KL}{1+\delta}$, where
$KL$ is the on-momentum focusing strength. In the high disruption beam 
most losses are caused by the low energy over-focusing. This happens
when the quadrupole off-momentum focal distance $\frac{1+\delta}{KL}$ 
becomes too small compared to the distance between quadrupoles. 
One way to reduce this effect is to make smaller the on-momentum 
$KL$ values. This can be done by increasing the number
of alternating gradient quadrupoles after IP. Since the total on-momentum
focusing in the quadrupole system has to remain about the same 
(to focus to the secondary IP), the individual quadrupole $KL$ values 
are reduced with more quads. As a result,
the low energy particles experience less deflections in each quadrupole
and less betatron amplitude in the system. The disadvantages of
this method are the longer quadrupole system, the increased on-momentum
beam size and larger aperture. 

In this study, we added one more quadrupole to make a six-quadrupole
system after IP. Due to the larger aperture and additional
quadrupole, the length of this focusing system increased
from the current 25.2~m to 45.8~m. The quadrupole parameters were optimized
to provide at least $5\sigma_{x}$/$10\sigma_{y}$ disrupted beam acceptance
in the quadrupoles within the energy range of $\delta$ from 0 to -70\%. 
The resultant apertures vary from $a_{p}=23$~mm in 
the first quadrupole to 69~mm in the sixth quadrupole (compare to
$a_{p}=10$ to 32~mm in the current lattice).

\begin{figure}[b!]
\includegraphics{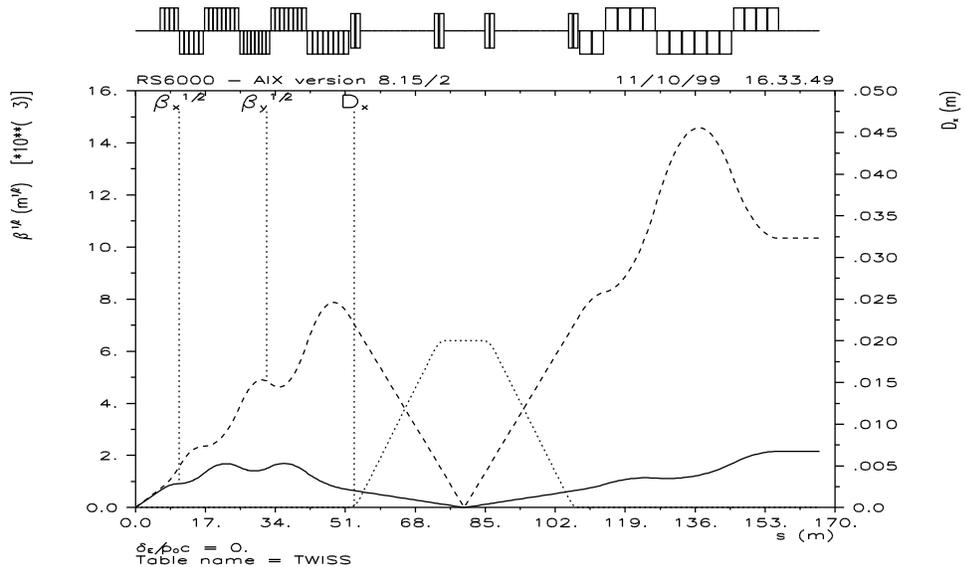}
\vspace{76mm}
\caption{Lattice option for the high disruption beam.}
\label{ext:lat-high}
\vspace{0mm}
\end{figure}

The described lattice option for the high disruption beam is shown in
Fig.~\ref{ext:lat-high}. The parameters of the last quadruplet were also 
optimized for the maximum acceptance. The apertures of these quadrupoles
range from $a_{p}=114$ to 157~mm and were defined by the $\pm 1$~mrad
IP divergence of the beamstrahlung photon beam. In the chicane, the
bend $x/y$-apertures were increased from $\pm 202/50$~mm 
to $\pm 250/108$~mm. For realistic
design, the long quadrupoles were made of a number of short quadrupoles
with gaps between them.

\begin{figure}[p]
\includegraphics{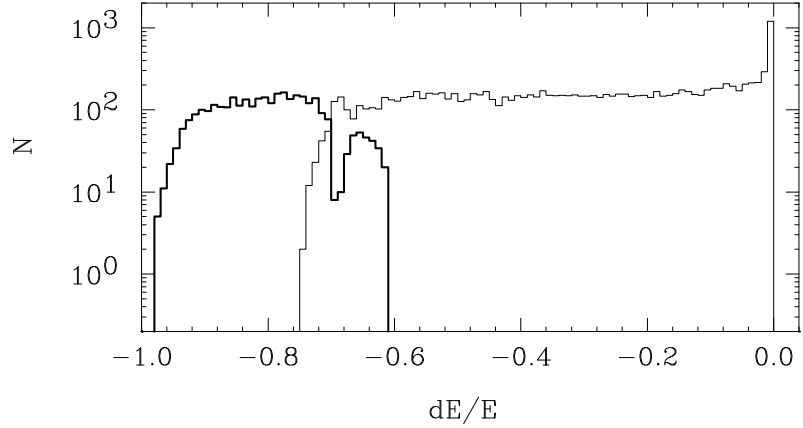}
\vspace{56mm}
\caption{Energy distribution for the lost (thick line) and survived
(thin) high disruption beam particles in the modified lattice
(15,000~particles).}
\label{ext:de-lost-high-dump13}
\vspace{0mm}
\end{figure}

Tracking simulation with the high disruption beam showed
that the total beam loss in the modified lattice is still very high:
21.2\% particles versus 28.2\% in the original lattice. Despite
the larger quadrupole apertures after IP, about 14.7\% particles
are lost in the six-quad system. The energy distribution of the lost
and survived particles is shown in Fig.~\ref{ext:de-lost-high-dump13}. 
Clearly, all of the beam losses occur in the low energy tail at
$\delta$ below -60\%. The analysis of the beam distribution at IP
showed that the highest energy lost particles ($\delta \approx -60$\%) 
had also large horizontal angles at IP, while the particles with small
IP angles survived up to $\delta \approx -75$\%.

\begin{figure}[p]
\includegraphics{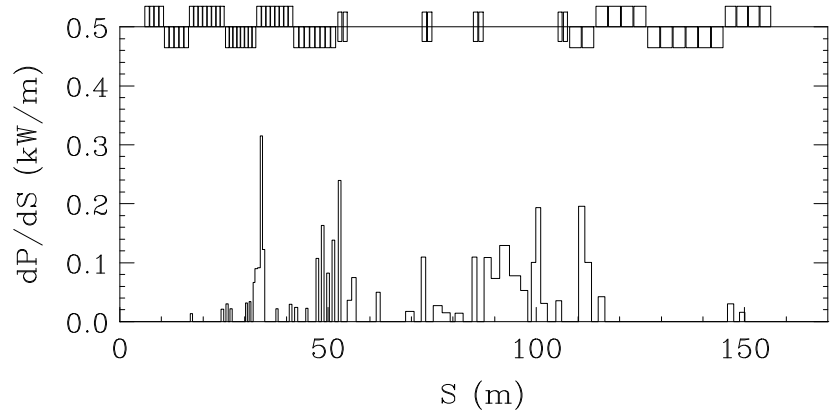}
\vspace{54mm}
\caption{Power loss distribution for the nominal disruption beam
in the modified lattice (NLC~1~TeV~case~A).}
\label{ext:power-nom-dump13}
\vspace{0mm}
\end{figure}

\begin{figure}[p]
\includegraphics{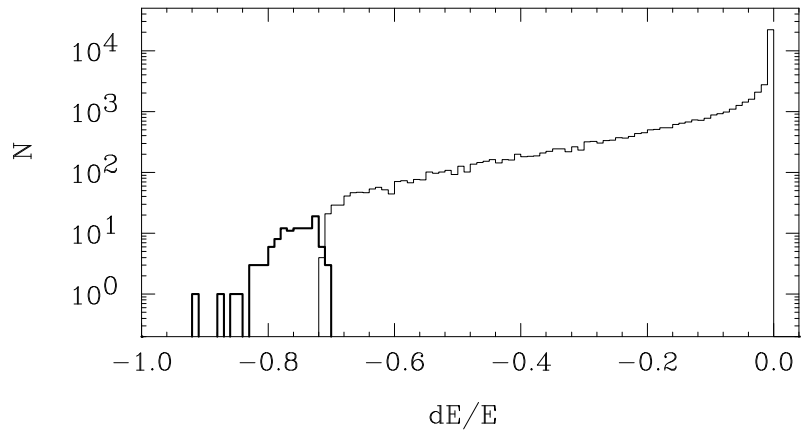}
\vspace{56mm}
\caption{Energy distribution for the lost (thick line) and survived
(thin) nominal disruption beam particles in the modified lattice 
(NLC~1~TeV~case~A; 50,000~particles).}
\label{ext:de-lost-nom-dump13}
\vspace{0mm}
\end{figure}

For comparison, we tracked the nominal disruption beam in this
modified lattice. The total beam power loss reduced to 3.9~kW
(4.8~kW in the original lattice), and the number
of lost particles reduced from 0.25\% to 0.23\%. The power loss distribution
in the modified lattice is shown in Fig.~\ref{ext:power-nom-dump13} and the
energy distribution of the lost and survived particles is shown
in Fig.~\ref{ext:de-lost-nom-dump13}. Note that increased bandwidth
in Fig.~\ref{ext:de-lost-nom-dump13} compared to
Fig.~\ref{ext:de-lost-high-dump13} is due to smaller IP angular
divergence in the nominal disruption beam.

Since the particle deflections in the quadrupoles increase at low
energies rather fast ($\sim 1/energy$), very large magnets would be needed
for a significant improvement of the chromatic acceptance with the
high disruption beam. Further discussions are needed to decide if the
use of larger magnets is acceptable.

\subsection{Sextupole Correction}

In accelerators sextupoles are typically used to compensate linear 
and some high order deviations in quadrupole focusing caused by
energy errors. The correction requires non-zero dispersion at 
the sextupoles to generate compensating quadrupole effect on the orbit
of the off-momentum particles. For a small energy spread, the focusing 
in quadrupoles and sextupoles depends linearly on $\delta$, therefore 
the correction works for all particles in the bunch.
In the case of the high disruption beam with a very broad energy spread,
the focusing and trajectories change non-linearly at large 
$|\delta|$ 
and the sextupole effect may vary significantly for particles
with different energies.

To study the effect of sextupoles on high disruption beam loss,
we added a sextupole to each of the first six quadrupoles in the
modified extraction line lattice
and a horizontal bend at $s=2$~m from IP (4~m before the first quad)
to generate dispersion. For the maximum effect, we assumed that the 
sextupole field can be superimposed over full length of the quadrupoles.
Though a large dispersion helps reduce the sextupole field,
its amplitude is limited by the quadrupole aperture $a_p$ and desired 
energy acceptance. In our case, the bend at $s=2$~m generates spread in
off-momentum particle trajectories, so at the entrance of the first
quadrupole the horizontal orbit as a function of energy is
$x(\delta)=\eta\frac{\delta}{1+\delta}$, where $\eta$ is the
linear dispersion at this point. Applying
$|x(\delta)|<a_{p}$ for $a_{p}=23$~mm and $\delta>-92$\% results in the
maximum dispersion value of $|\eta|=2$~mm. This corresponds to
8.34~kG$\cdot$m field in the dispersion generating bend at 500~GeV.

Optimization of the sextupole strengths showed that this correction
does not significantly improve the low energy beam acceptance.
With the sextupole field effectively imposed over quadrupole length
and for the optimized values of the pole tip field in the six sextupoles
($B_{p}^{sext}=4,0.5,-9.5,13,-5,-1$~kG), the particle loss at the end
of the six-quadrupole system is reduced from 14.7\% to 11.3\%. This 
marginal reduction in the beam loss seems to not justify the complications
of an extra bend near IP, superimposed sextupole and quadrupole fields
and the need for dispersion cancellation after the sextupoles.

\subsection{Octupole Correction}

In a separate study, we investigated the effect of octupoles on the
beam loss. As in the case with sextupoles, we effectively superimposed
six octupoles over the six quadrupoles after IP. Since the
octupole field increases as a cube of a particle amplitude, we consider
that its most effect will be on the low energy particles experiencing
large deflections in the quads, and will not significantly affect the
rest of the beam. 

In quadrupoles the beam is focused or defocused either in
horizontal or vertical direction. Accordingly, the main losses of the
low energy particles are expected to be in $x$ and $y$-planes (see
Fig.~\ref{ext:theta-lost}). Octupoles also provide horizontal and vertical
focusing which can be used to compensate the quadrupole effect at large
amplitudes. Compared to the quadrupoles, a normal octupole provides 
simultaneous non-linear focusing along the $x$ and $y$-axes and 
defocusing along the $|x|=|y|$ lines, or vice versa.

In the quadrupole system the net $x$ and $y$
focusing is dominated by the horizontally (F) and vertically (D) focusing 
quads, respectively. Therefore, to reduce the net low energy over-focusing
in both planes one has to reduce the horizontal focusing in F-quads and
vertical focusing in D-quads. Since the octupoles simultaneously focus 
(or defocus) in the $x$ and $y$-planes, all six 
compensating octupoles need to be defocusing in order to reduce the 
quadrupole focusing at large amplitudes. 

\begin{figure}[h]
\includegraphics{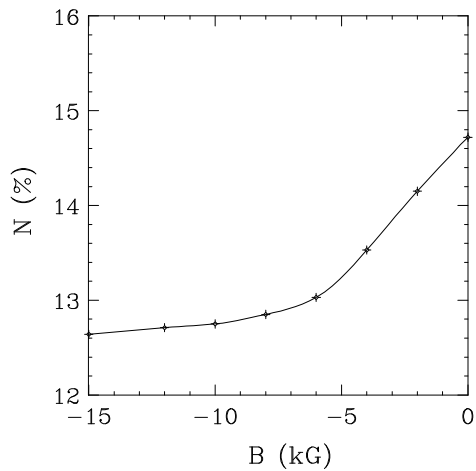}
\vspace{64mm}
\caption{High disruption beam loss $N$(\%) at the end 
of the six-quadrupole system versus octupole pole tip field.}
\label{ext:oct-lost}
\vspace{0mm}
\end{figure}

In simulations, we varied the individual octupole strengths and confirmed
that the six octupoles have to be defocusing for the most reduction of
the high disruption beam loss. The same aperture and effective 
length was used for the combined octupoles and quadrupoles. 
Fig.~\ref{ext:oct-lost} shows the particle loss
as a function of octupole pole tip field (same for the six 
octupoles). For realistic field values the particle loss at the end of
the six-quadrupole system reduces to 12.6\% from 14.7\% without octupoles.
This is rather small improvement and we do not yet consider 
this compensation as a practical solution.

\subsection{Low Energy Tail Separation}

An alternative way to avoid large beam loss in the extraction line
is to separate the low energy tail from the beam
and direct it to a different transport line. In this case, the
energy acceptance requirement can be relaxed for the main beam extraction
line and the transport line for the tail can be better adjusted for
the lower energy or designed to absorb the energy. Since the large
losses start in the quadrupoles after IP, the separation has to be
done before the beam enters the quads. 

\begin{figure}[b!]
\includegraphics{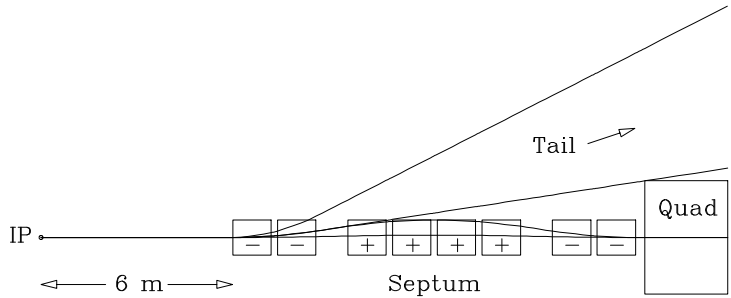}
\vspace{46mm}
\caption{Beam separation with chicane and septum bends.}
\label{ext:separ}
\vspace{0mm}
\end{figure}

Dipole bending naturally spreads the particle trajectories with
different energies. The low energy particles experience the largest
deviations from the on-momentum orbit, which are proportional to 
$\frac{\delta}{1+\delta}$. For the beam separation study, 
we introduced a symmetric horizontal chicane between 
$s=6$ and 18.6~m after IP and moved the first quadrupole
to $s=18.9$~m as shown in Fig.~\ref{ext:separ}. Each individual 
chicane bend is 1.2~m long and has 12~kG field at 500~GeV. 
The goal for the first two chicane bends was to generate a large
spread between trajectories of the low energy particles and
the core beam. Then the septum bends in the middle of chicane are used
to deflect back only the core beam particles
and allow the low energy tail travel unbent into a different
transport line. The dispersion at the end of the chicane is 
canceled, and if the energy spread in the beam is significantly
reduced after the tail separation, then a simpler quadrupole system, 
such as a doublet, can be used to focus to the secondary IP.

We used the beam separation condition which required that the separated tail
particles did not strike the face of the first quadrupole located at 
$s=18.9$~m. In other words, the septum bends would separate only the tail 
particles with deflection at the first quadrupole larger than the outer 
quadrupole radius. We estimated that the aperture
of the first quadrupole needs to be about $a_{p}\approx 60$~mm and the outer
radius of such a permanent magnet with 12~kG pole tip field can be on the
order of 150~mm. 

\begin{figure}[t]
\includegraphics{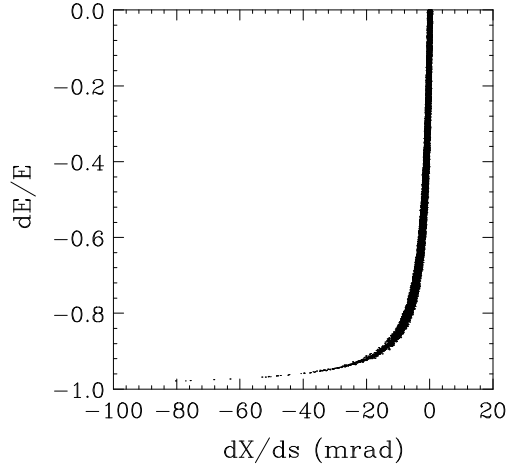}
\vspace{64mm}
\caption{Horizontal angular distribution vs. energy at the
separating septum bend.}
\label{ext:de-px-separ}
\vspace{0mm}
\end{figure}

\begin{figure}[t]
\includegraphics{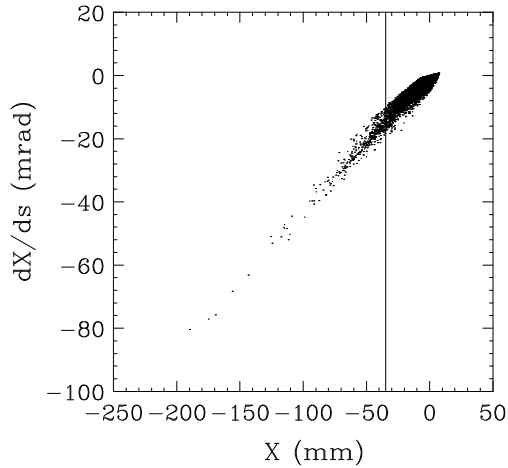}
\vspace{64mm}
\caption{Horizontal phase space distribution of the high disruption
beam at the separating septum bend. The vertical line separates the tail 
(on the left) from the core.}
\label{ext:x-px-separ}
\vspace{0mm}
\end{figure}

\begin{figure}[t]
\includegraphics{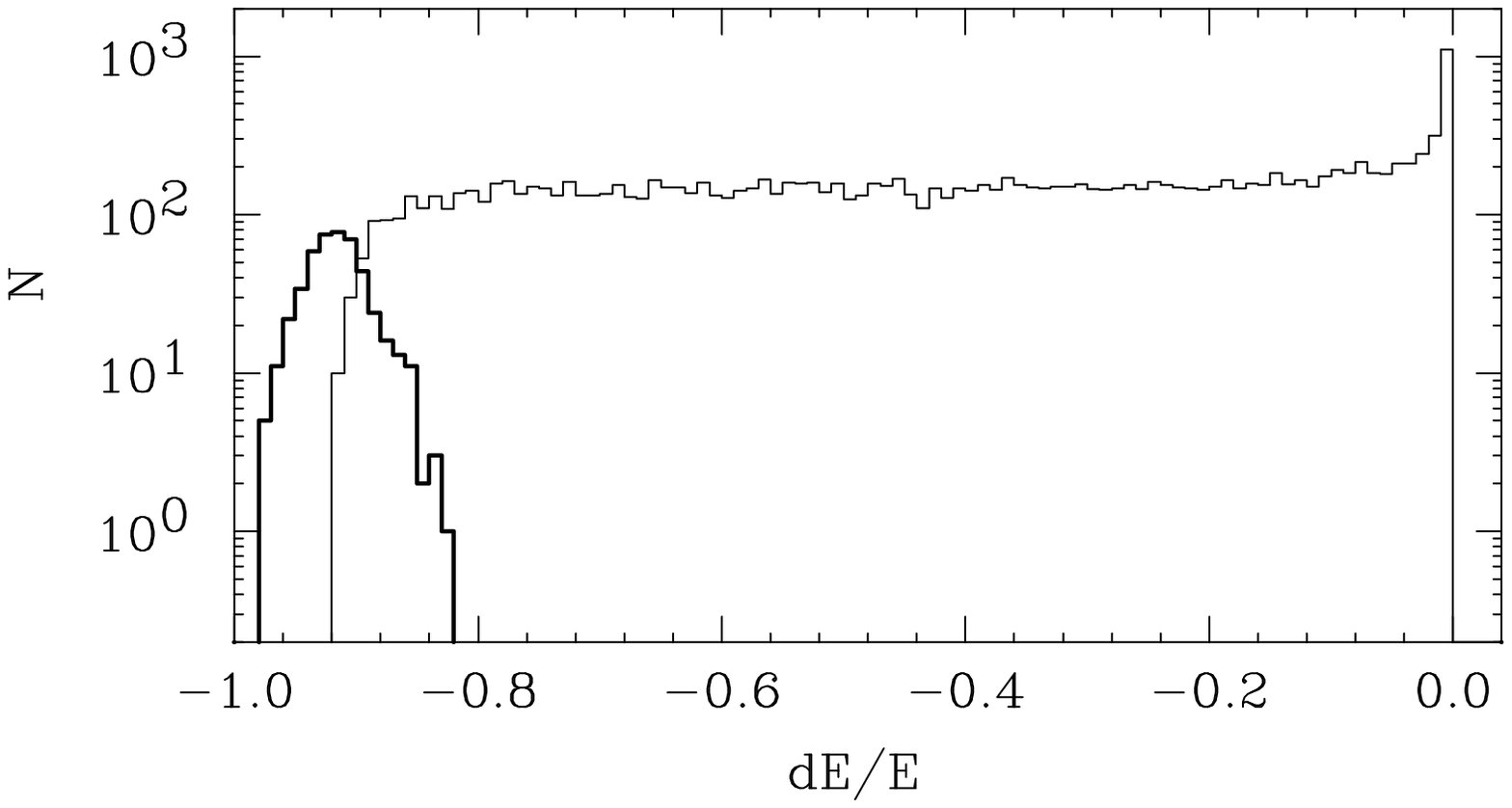}
\vspace{56mm}
\caption{Energy distribution for the tail (thick line) and the core
(thin) high disruption beams at the separating septum bend 
(15,000~particles).}
\label{ext:de-separ}
\vspace{0mm}
\end{figure}

The high disruption beam distribution was tracked through the first
two deflecting
bends, and Fig.~\ref{ext:de-px-separ} shows the horizontal angular
divergence versus energy at the entrance to the first separating septum
bend at $s=9.6$~m. The horizontal phase space at this point is shown in
Fig.~\ref{ext:x-px-separ}, where the vertical line at
$x=-34.8$~mm separates the beam core and the tail. Counting the
particles shows that only 3.1\% particles are in the tail.
The energy distributions of the core and tail beams are shown in
Fig.~\ref{ext:de-separ}.

Since the number of low energy particles in the core beam is still
large after the separation, we conclude that the above scheme is not 
very effective. For this reason we did not proceed to study other
important issues such as the design and efficiency of septum bends 
and the transport line for the low energy tail.

\section{Summary}

We investigated various methods to reduce the beam loss in the NLC
extraction line caused by a large number of very low energy particles in the
high disruption beam option. The study included the use of larger
quadrupole aperture, weaker quadrupole focusing, sextupole compensation,
octupole focusing at large amplitudes, and beam tail separation with
chicane septum bends. We did not find significant reduction of the beam
loss with the above methods. Further studies and new ideas are needed
to find a practical solution for the high disruption scenario.

\end{document}